# Shubnikov-de Haas (SdH) Oscillation in Self-Flux Grown Rhombohedral Single-Crystalline Bismuth

Yogesh Kumar [a, b], Prince Sharma [a, b], N. K. Karn [a, b], V.P.S. Awana [a, b, *]

[a] *CSIR-National Physical Laboratory, Dr. K.S. Krishnan Marg, New Delhi-110012, India*
[b] *Academy of Scientific and Innovative Research (AcSIR), Ghaziabad, U.P. 201002, India*

ABSTRACT

The historic de Haas-van Alphen effect observed in the late 1950s in CSIR-NPL by J.S. Dhillon and D. Shoenberg in pure bismuth, and zinc metal is revisited in this article through a single crystalline phase of bismuth crystal, which is observed in terms of resistivity as predicted by Shubnikov–de Haas (SdH) oscillations. The single crystal of bismuth is grown through solid-state reaction under an optimized heat treatment whose purity and structural phase are confirmed through XRD, SEM, and EDAX. The transport properties of single-crystal show the presence of SdH oscillations from a temperature range of 2K to 10K. The occurrence of oscillations in the transverse magnetic field confirms the presence of the Fermi surface. Landau level (LL) fan diagram reveals the presence of topological surface states and the Berry's phase confirmation from temperature-dependent SdH oscillations, concluding the presence of a noble topological phase of bismuth exhibiting SdH oscillations.

**Keywords**: Single crystal bismuth, Berry phase, SdH oscillations, Topological bismuth, LL fan diagram, Dingle temperature.

INTRODUCTION

Bismuth is an all-time fascinating material for condensed matter physicists. Starting from the late 1778 discovery of large anomalous diamagnetism to Seebeck effect in 1821, Nernst effect 1886, Kapitza's law of magnetoresistance 1928, De Haas Van Alphen effect 1930, Shubnikov-de Haas effect 1930, cyclotron resonance 1955, oscillatory magnetostriction 1963, topology 2006, and anomalous quantum oscillation in longitudinal magneto thermoelectric power in 2018, have resolved a lot of vague uncertainties of science [1-12]. The historical phenomenon of the de Haas-van Alphen effect was experimentally observed in the late 1930s by WJ De Haas et al. [6], and it was revisited in CSIR-NPL by J.S. Dhillon and D. Shoenberg in the 1950s [9]. J.S. Dhillon et al. observed de Haas-van Alphen effect in pure bismuth and zinc metal through torque method at 4.19K and 1.5K between about 1.5 and 32kG [9]. In both these experiments by W.J. De Haas et al. [6] and J.S. Dhillon et al [9], oscillations in magnetization with respect to the field were observed in bismuth. Moreover, J.S. Dhillon et al. considered various elements such as Zn, Bi, Pb, Ag, Au, and Cu for their study and only Bi and Zn were found to have oscillations in their magnetization analysis [9]. Furthermore, at that time, J.S. Dhillon et al. concluded that these oscillations were dependent on temperature, and related to its temperature dependence of steady susceptibility [9]. In 1962, Shubnikov-de Haas came up with the idea of finding oscillations in resistivity, same as being observed in magnetization as a function of the magnetic field [7]. This time, Shubnikov-de Haas observed oscillations in the transverse magnetoresistance of bismuth metal and predicted the Fermi surface of bismuth in three and four carriers models [7]. Further, these oscillations were studied in various Sn, Pb, and Sb doped Bi along with nanoribbons, nanowires, electron deposited Bi thin films, single crystals, and tin ionized Bi [9,13–30].

Later, in 2005, the term topology changed the scenario of condensed matter physics as Charles Kane and Eugene Mele theoretically predicted the existence of topological insulators (TI) [31–34]. TI has a unique property of insulating at the bulk, while the surface shows the exotic state which is conducting. The electronic wave function of the charge carriers is dependent on the geometry as it changes from bulk to surface [35–37]. The surface states (SS) makes the system relentless to the nonmagnetic doping as the SS are protected by time-reversal symmetry (TRS). Such a trifling character provides numerous novel applications as the topological states are robust to nonmagnetic perturbation and show dissipation-less spin current transport. These applications include spintronic, thermo-electric, magnetic memory storage, magneto-electric devices, next-generation batteries, THz generators, transistors, photodetectors, and sensors applications [38–40]. Among these TIs, Bismuth is found to have high order intrinsic topology [5,8,31,34,36,41,42]. Thus, the bismuth metal has been fascinating for condensed matter physicists from the late 1950s to till date. This article revisited the SdH oscillation in the pure bismuth single crystal. The amplitude of SdH oscillations is found to be dependent on temperature, which is modelled to show its topological phases. While the temperature and field dependence of SdH oscillations reveals the information about 2D charge carriers. The short article on low temperature magneto transport of rhombohedral Bi crystal beautifully clubs together the topology and the Shubnikov-de Haas quantum oscillations.

EXPERIMENTAL DETAILS:

High purity (99.999%) bismuth (Bi) powder was taken to grow Bi single crystal using the self-flux method via solid-state reaction route [43]. The Bi powder was pelletized into a rectangular pellet using a hydraulic press and vacuum-sealed into a quartz tube with a pressure of $10^{-5}$ mbar. The ampoule was put into an auto-programmable furnace, where the sample was heated from room temperature to 650°C at a rate of 40°C/h and kept hold for 8h. For single crystalline growth, the molten metal was then cooled down to 250°C with a slow cooling rate of 3°C/h and kept hold for 95h at 250°C. Further, the ampoule was cooled down to 245°C and kept there for another 73h; finally, it was allowed to cool down to room temperature with a cooling rate of 120°C/h, see for more details ref. 43. Figure 1 shows the detailed heat treatment, and the inset image shows the obtained silvery shiny Bi crystal of size ≈ 1cm×0.5cm. For characterization, i.e., structural and transport properties, thin flakes were mechanically cleaved from an obtained single crystal. The table top Rigaku Miniflex II (X-ray diffractometer with radiation Cu-K$_\alpha$ (λ = 1.5406Å)) was used to record diffraction spectra at room temperature, while the powder XRD pattern was Rietveld refined using FullProf software, and its crystal structure was extracted through VESTA software. The scanning electron microscopy (SEM) equipped with energy-dispersive X-ray spectroscopy (EDAX) was performed to study its morphology and elemental composition. The vibrations modes were recorded by a Renishaw Raman spectrometer equipped with a 514 nm laser. The transport measurements were performed using the quantum design physical property measurement system (PPMS) up to 14Tesla and down to 2K.

RESULTS AND DISCUSSION:

To investigate the phase purity of synthesized Bi crystal, the mechanically cleaved flakes were crushed into powder, and X-ray diffraction (XRD) spectra of the powder were recorded at room temperature. Figure 2(a) depicts the Rietveld refinement of powder XRD spectra in a 2θ range from 20-80°. The fitting of all diffraction peaks confirms that the grown crystal is in a single phase. The synthesized crystal belongs to a rhombohedral crystal structure with $R\bar{3}m$ (166) space group having refined lattice parameters a = b = 4.546(1)Å and c = 11.859(6)Å. Further, to investigate the growth direction of grown crystal, XRD spectra were recorded on mechanically cleaved flake, which reveals that Bi crystallized along (00$l$) diffraction plane with $l$ = 3n (n = 1,2,3…) as shown in figure 2(b). An additional peak at 2θ = 48.82° is also observed, which may be due to the misalignment of the crystal, as mentioned in the literature [43]. The Rietveld refined parameters were used to draw Bi unit cells using VESTA software, as shown in the inset of Fig. 2(b).

The vibrational modes of single-crystalline flake were recorded at room temperature, and Fig.3 shows the observed Raman modes of studied bismuth crystal. Two peaks were observed at 69.84 and 96.91cm$^{-1}$, which corresponds to $E_g$ and $A_{1g}$ vibrational modes respectively. The observed peaks are in good agreement with previous reports [44–46]. The surface morphology of as-grown crystal was studied by performing SEM on cleaved Bi flake and is shown in the inset of Fig.3. The layered growth along one direction and absence of any grain boundaries reveals the crystalline nature of synthesized Bi, which is in accordance with XRD spectra in Fig. 2(a). Another inset represents the EDAX, which shows the absence of any impurity element in as-grown Bi crystal.

The transport properties are investigated using Quantum design PPMS. Figure 4(a) represents the temperature dependence of longitudinal resistivity in zero magnetic fields applied perpendicular to the sample. The four probe arrangement is used to measure resistivity of sample and the applied field is perpendicular to sample surface, as illustrated in lower inset schematic of fig 4(a). It is observed that with decreasing temperature from 300K to 2K, the resistivity is monotonically decreased from 1.35mΩ-cm to 0.21mΩ-cm, which signifies the metallic nature of the grown crystal. The residual resistivity ratio [RRR = $\rho(300K)/\rho(2K)$] is found to be 6.43, which corresponds to high crystalline nature, and this value is in accordance with previous reports on TI single crystals [47,48]. The left inset of Fig. 4(a) represents the electrical resistivity behaviour as a function of transverse magnetic field, ranging from zero to 14Tesla at different temperatures, varying from 2K to 100K.

To investigate the magneto transport, the magnetoresistance (MR) has been determined using the formula MR% = $(\rho(H) – \rho(0))/\rho(H) \times 100\%$ where $\rho(H)$ and $\rho(0)$ are resistivity at applied and zero magnetic fields respectively. Figure 4(b) shows the transverse MR% of Bi crystal in a magnetic field up to 14Tesla at different temperatures. The plotted MR% is calculated by averaging the measured MR data in both positive and negative field directions. It is observed that MR varies linearly with the magnetic field at all measured temperatures and is non-saturating up to 14Tesla. For temperature 2K, the value of MR% is ≈ 1700% at the highest measured magnetic field 14Tesla which further decreased to ≈ 1000% as the temperature is increased to 100K.

In MR measurements, the signature of Shubnikov–de Haas (SdH) oscillations has been observed at low temperatures. To further explore the observed SdH effect in Bi crystal, the resistance versus applied magnetic field measurements have been performed by varying the temperature between 2K to 10K. For clarity and better understanding, a derivative of resistance with respect to the applied field at different temperatures is shown in Fig. 5(a). The oscillatory behaviour of dR/dH displays the periodic variation of maxima or minima against the inverse magnetic field, which is due to the presence of the SdH effect. Also, the dependence of maxima/minima positions on transverse magnetic fields suggests that SdH oscillations do possibly originate from 2D surface states [49]. The occurrence of oscillations in the transverse magnetic field confirms the presence of the Fermi surface, and its cross-section area depends on the frequency of these oscillations. The fast Fourier transform (FFT) of dR/dH vs. 1/H has been performed at respective temperatures to investigate these oscillations, as shown in the inset of Fig. 5(a). The occurrence of a single frequency peak at ≈ 18.92Tesla indicates the existence of a single pocket near the Fermi surface. According to Onsager relation, the SdH oscillation frequency (F) is proportional to the cross-section area ($A_F$) of the Fermi surface by relation F = ($\hbar/2\pi e$)$A_F$. Here, $\hbar$ and e are Plank's constant divided by $2\pi$ and electronic charge respectively, the calculated cross-sectional area of Fermi surface corresponding to frequency 18.92Tesla which comes out to be 180.59 × $10^{15}$m$^{-2}$. Assuming a circular cross-section of the Fermi surface, the Fermi wave vector ($k_F$) is described as $k_F = (A_F/\pi)^{1/2}$. The obtained value of the Fermi wave vector corresponding to $A_F$ is found to be 0.024 Å$^{-1}$, which is consistent with the literature [49]. Also, the 2D surface carrier density ($n_{2D}$) can be calculated from the Fermi wave vector by using relation $n_{2D} = k_F^2/4\pi$, the obtained value of $n_{2D}$ is found to be $4.57 \times 10^{11}$ cm$^{-2}$ which is comparable and in accordance with previous reports on topological insulators [20,50,51].

In order to investigate the topological nature of electrons, the occurrence of Berry's phase has been extracted from quantum oscillations data through the Landau level (LL) fan diagram. In the LL fan diagram, the Landau index (n) is plotted as a function of the inverse magnetic field, as shown in the main panel of Fig. 5(b). Here the integral LL index has been assigned to the minima of quantum oscillations in the dR/dH vs. 1/H plot for the entire field range at 2K, and the LL index is increased by 1 between adjacent minima [50,52]. According to Lifshitz-Onsager quantization rule, the LL index n is inversely proportional to the magnetic field, which can be expressed as $A_F\hbar/eH = 2\pi(n + 1/2 + \beta)$, here $2\pi\beta$ is Berry's phase [53,54]. In the LL fan diagram, the solid line represents the linear fitting of data points, and the extrapolation of this line gives the finite intercept of -0.531, which corresponds to 1.062π Berry's phase. The obtained value of Berry's phase is ≈ π, which confirms the topological nature of SdH oscillations arising from surface states and is in agreement with the literature on various topological insulators [55,56].

In order to extract more information about 2D charge carriers of SdH oscillations, the temperature dependence of oscillation amplitude has been investigated. It is observed that the oscillation amplitude decreases as the temperature is increased from 2K to 10K, as shown in Fig. 5(a). According to Lifshitz-Kosevich (LK) theory, the effective mass (m$^*$) of surface charge carriers can be calculated from the temperature-dependent oscillation amplitude by using the relation $\Delta R(T)/R(0) = \lambda(T)/\sinh(\lambda(T))$. In this expression, the $\lambda(T)$ is the thermal

7 | P a g e

factor, which is given as λ(T) = $2\pi^2 k_B T m^*/\hbar eH$, $k_B$ is Boltzmann's constant, $m^*$ is effective mass, T is the temperature, and e is electronic charge [57]. The extracted effective mass $m^*$ is $0.23 m_e$ (where $m_e$ is the mass of a free electron) from the theoretical fitting of oscillation amplitudes by using the LK equation as shown in the upper inset of Fig. 5(b). The obtained value is consistent with previous reports for topological insulators [58,59]. The corresponding Fermi velocity, $v_F = \hbar k_F/m^*$ and Fermi level, $E_F = m^* v_F^2$ are $7.6 \times 10^5$ m/s and 757.9 meV respectively for Bi single crystal. Further, the Dingle temperature can be extracted by studying the effect of field on amplitude of quantum oscillations at fixed temperature. The lower inset of Fig. 5(b) represents the Dingle plot for Bi crystal i.e., variation of lnD as a function of 1/H. The term D is expressed as, D = AHsinh(αT/H) where, A is the oscillation amplitude at particular 1/H value determined by averaging from adjacent peak/valleys and the factor α = $2\pi^2 k_B m^*/\hbar e$. The Dingle temperature ($T_D$) can be extracted from the slope of linear fitting of lnD vs 1/H data points. The solid line in lower inset of Fig. 5(b) represents the fitted linear curve and the obtained value of $T_D$ = 2.8K. The scattering time of charge carriers or carrier life time (τ) is found to be $4.2 \times 10^{-13}$s by using the relation τ = $\hbar/2\pi k_B T_D$. The corresponding carrier mobility, μ = $e\tau/m^*$ and mean free path, $l = v_F \tau$ of carriers are estimated to be 3289.9 cm$^2$ V$^{-1}$ s$^{-1}$ and 326.2 nm respectively. Thus, this study comprehensive investigates various intrinsic parameters of single crystalline bismuth which ensure the topological nature of self-flux grown single crystalline bismuth.

CONCLUSION:

Quantum SdH oscillations are seen at low T (below 10K) and higher fields (up to 14Tesla) in transverse magnetic field for studied rhombohedral Bi single crystal. The LL fan diagram shows a finite intercept of -0.501, which corresponds to π Berry's phase and concludes the topological nature of SdH oscillations arising from surface states. Along with the topological confirmation, various other parameters are also calculated, which includes effective mass m* of $0.23m_e$, Fermi velocity of $7.6 \times 10^5$ m/s, Fermi level of 757.9 meV, scattering time of charge carriers of $4.2 \times 10^{-13}$s, the carrier mobility of 3289.9 cm$^2$ V$^{-1}$ s$^{-1}$, and mean free path of 326.2 nm, respectively. This study investigates the intrinsic topological nature of bismuth using transport property analysis and predicts the application of bismuth in spintronics and quantum computing field.

ACKNOWLEDGMENT:

The authors would like to thank Director NPL for his keen interest and encouragement. Yogesh Kumar and Prince Sharma would like to thank CSIR and UGC for their fellowship support. The authors are also thankful to AcSIR for Ph.D. registration. Dr. J. S. Tawale is also acknowledged for performing SEM images.

AUTHOR INFORMATION

Yogesh Kumar
Electrical and electronics metrology division
CSIR-NPL, India.
Email-Id- yogesh30394@yahoo.com

Prince Sharma
Electrical and electronics metrology division
CSIR-NPL, India.
Email-Id- sharmapvats8@gmail.com

Navneet Karn
Electrical and electronics metrology division
CSIR-NPL, India.
Email-Id- navneeetkarn4@gmail.com

Corresponding Author
Dr. V. P. S. Awana
Electrical and electronics metrology division
CSIR-NPL, India.
E-mail: awana@nplindia.org
Ph. +91-11-45609357, Fax-+91-11-45609310
Homepage: awanavps.webs.com

FIGURE CAPTIONS:

Figure 1: Heat treatment schematic followed to grow Bi single crystal and inset shows the obtained silvery shiny single crystal.

Figure 2: (a) Rietveld refined powder XRD spectra of as grown Bi. (b) XRD spectra of mechanically cleaved Bi flake and inset shows the unit cell indicating Bi atoms.

Figure 3: Raman spectrum of Bi crystal, inset represents SEM image and atomic percentage of grown Bi crystal.

Figure 4: (a) Variation of electrical resistivity as a function of temperature from 300K to 2K, upper inset represents the dependence of resistivity on magnetic field up to 14 Tesla at different temperatures; lower inset shows the schematic of four probe resistivity measurement. (b) Magnetoresistance of Bi crystal at different temperatures.

Figure 5: (a) Behaviour of dR/dH of Bi crystal as a function of magnetic field at different temperatures, inset shows the temperature dependence of FFT spectrum of oscillations. (b) Landau level fan diagram, upper left inset shows oscillation amplitude as a function of temperature fitted with L-K formula and lower right inset is Dingle plot for Bi crystal.

Fig. 1

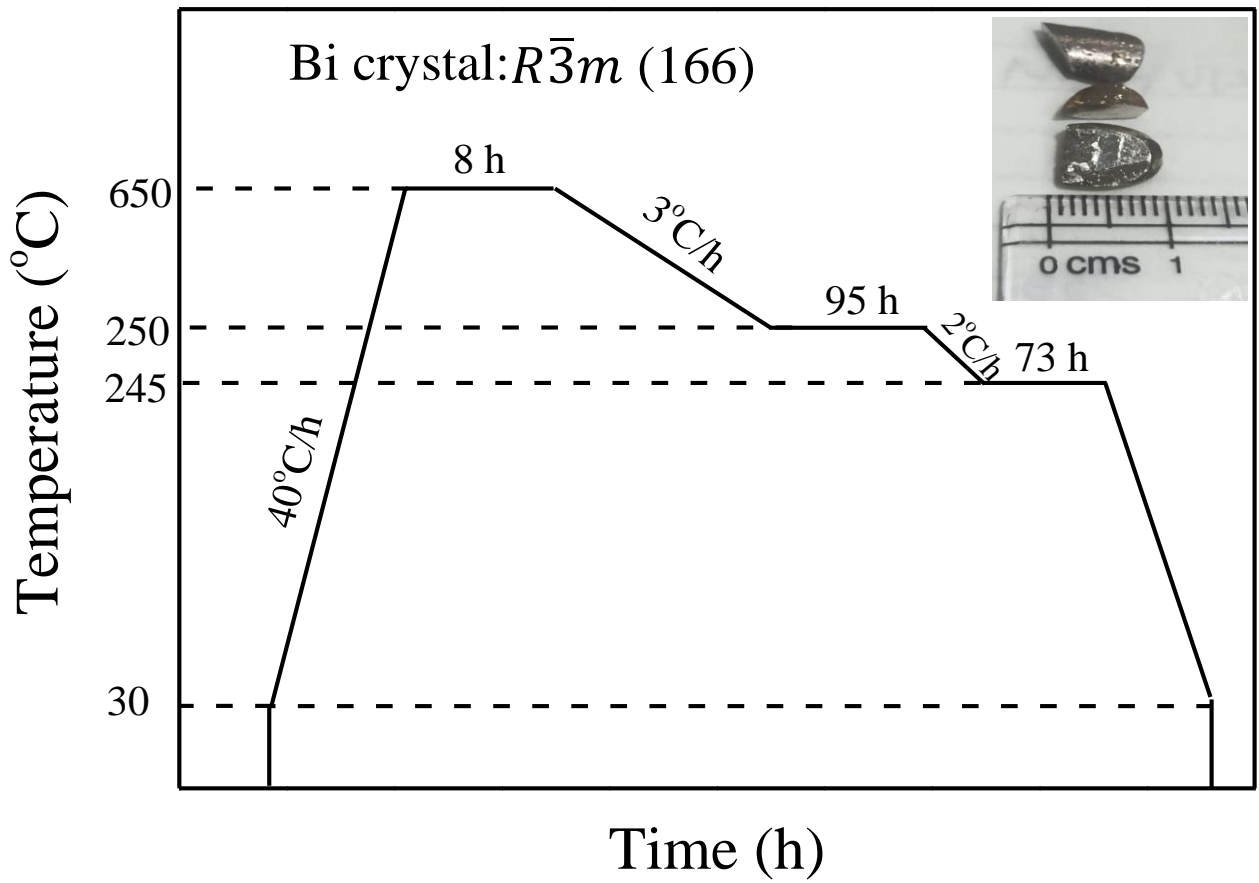

Fig. 2 (a)

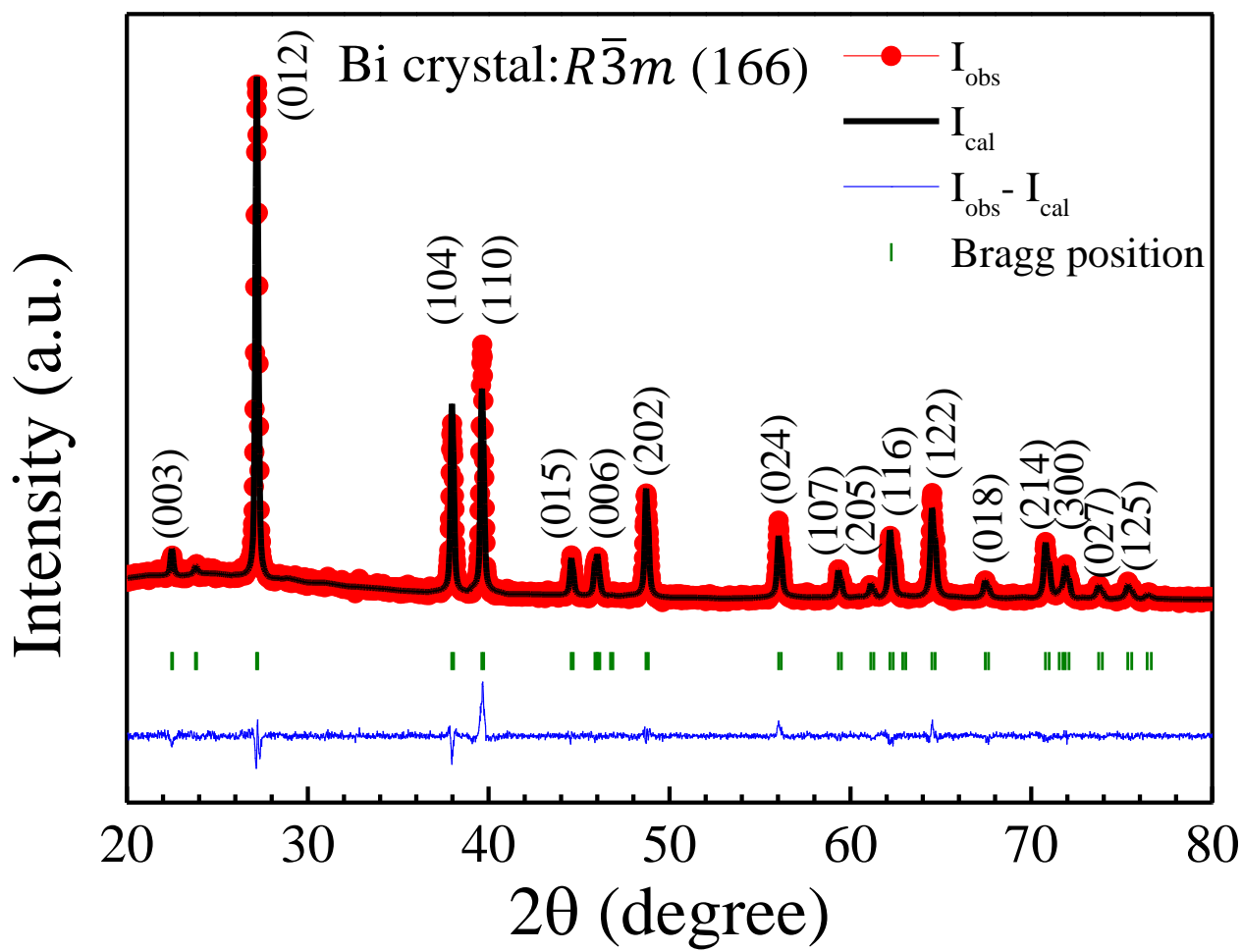

Fig. 2 (b)

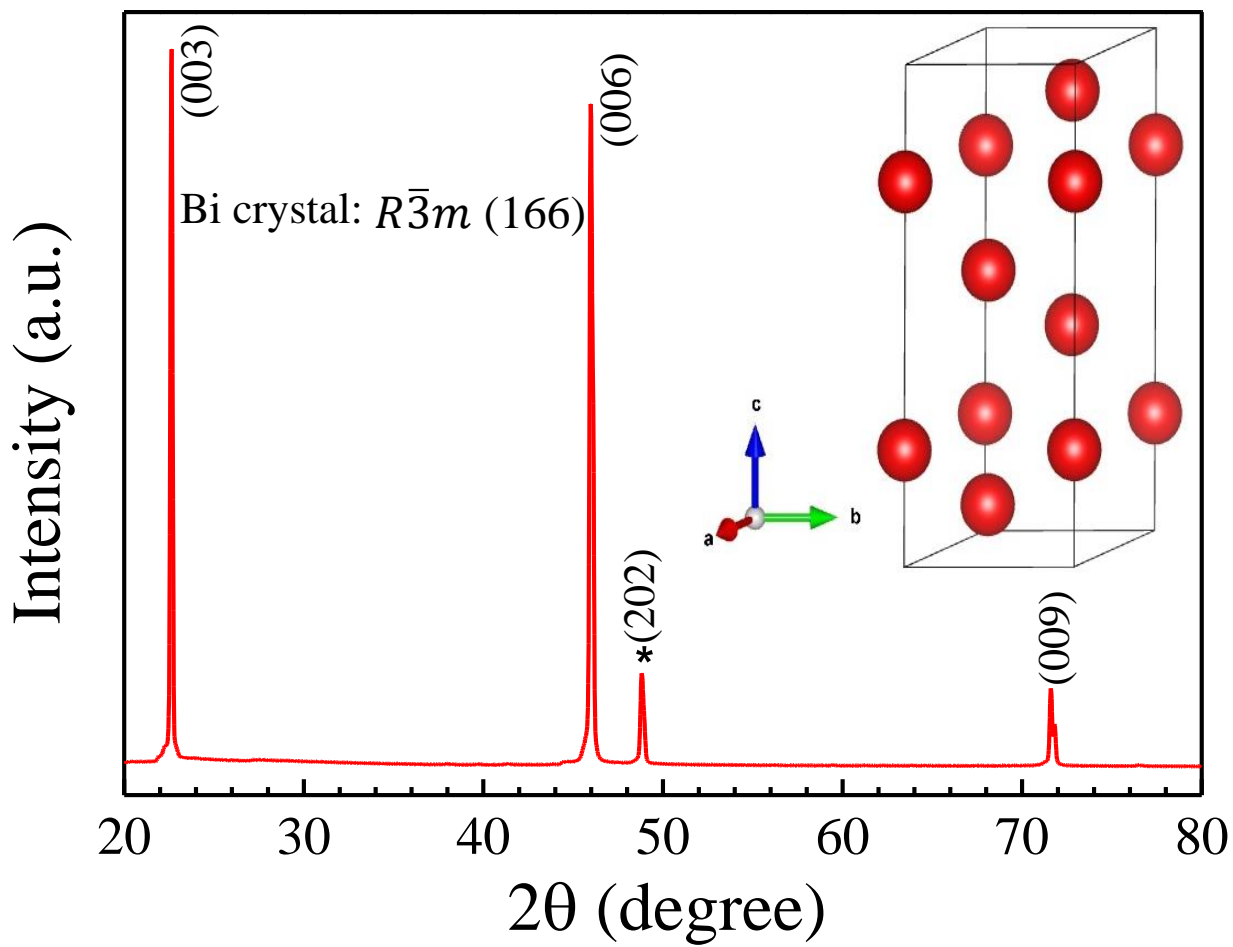

Fig. 3

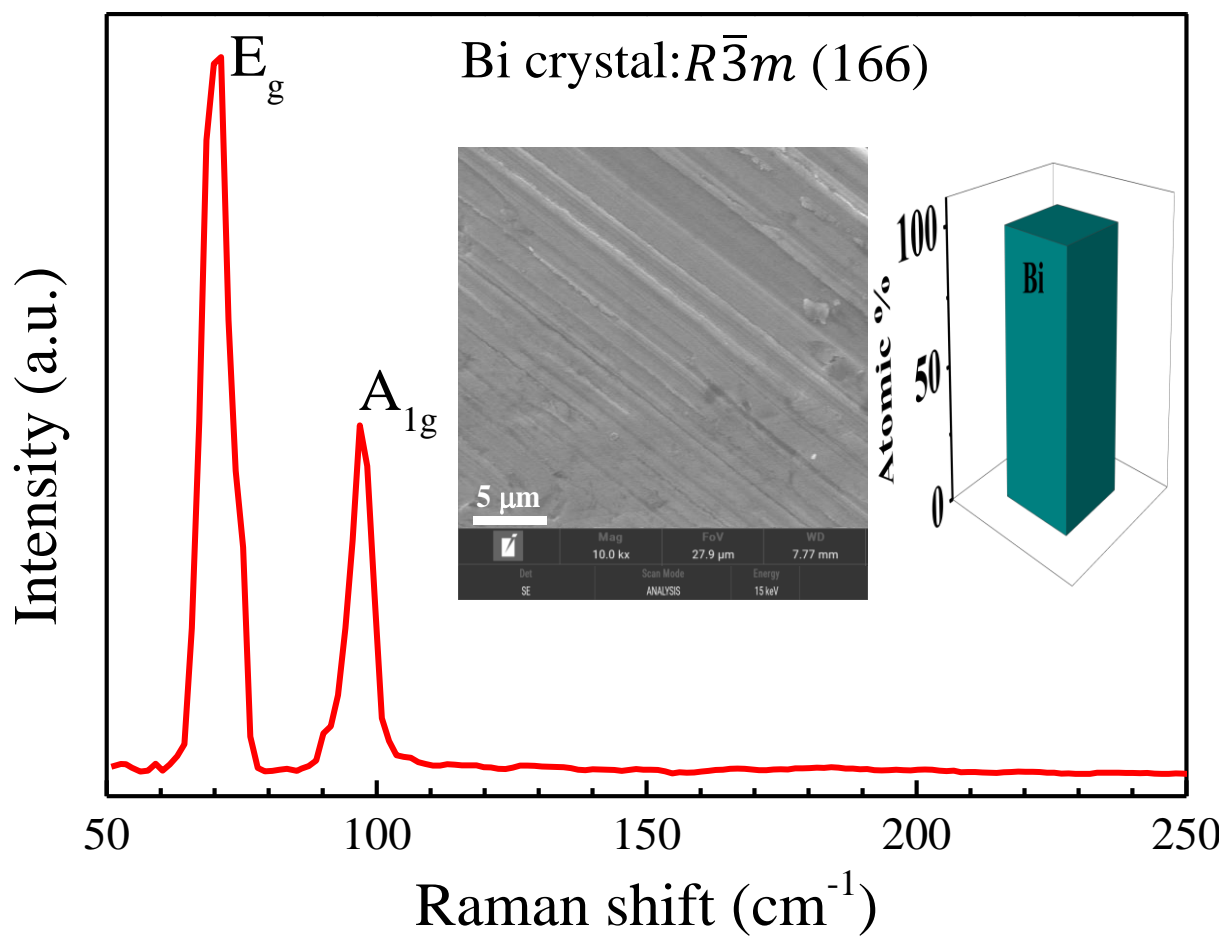

Fig. 4(a)

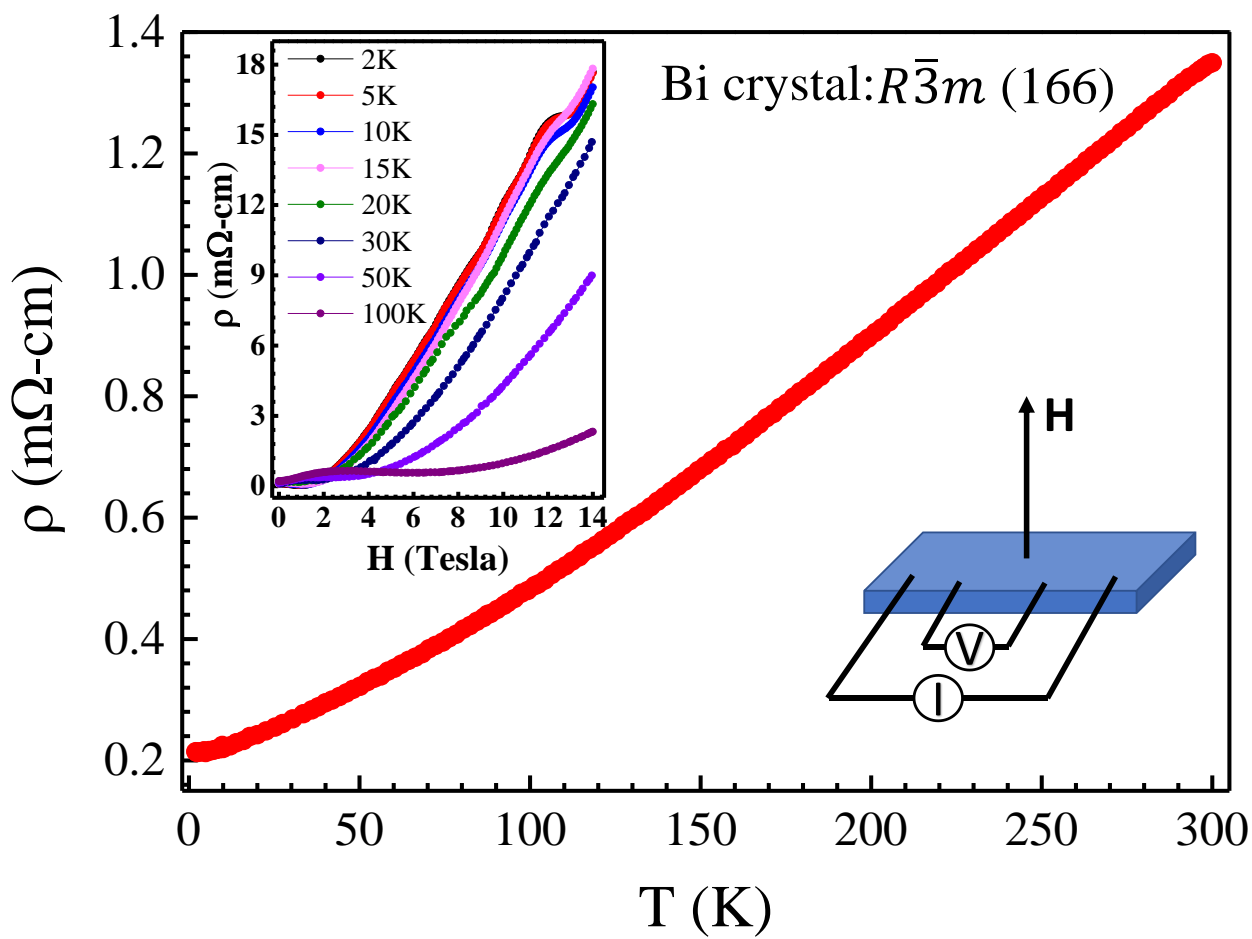

Fig. 4(b)

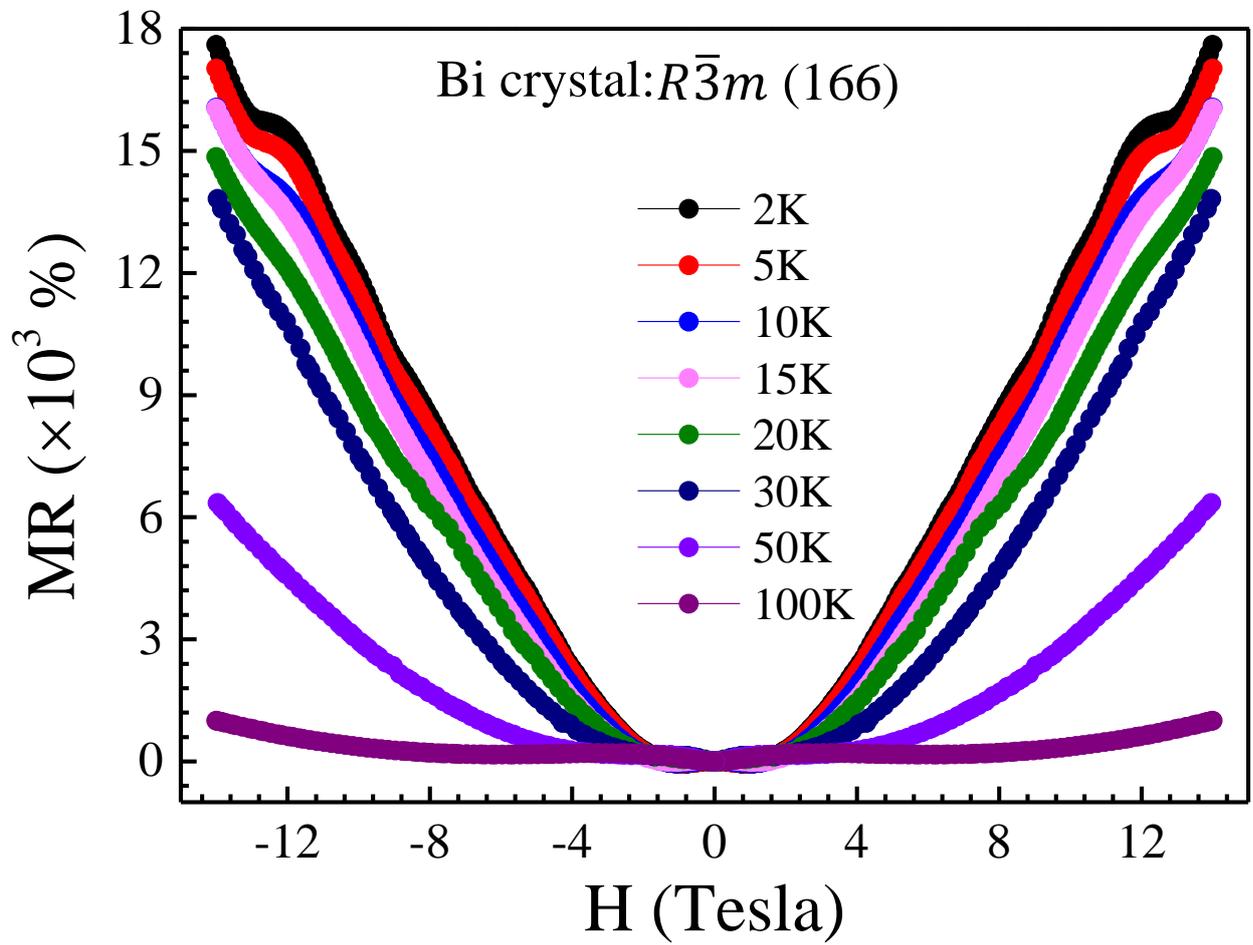

Fig. 5(a)

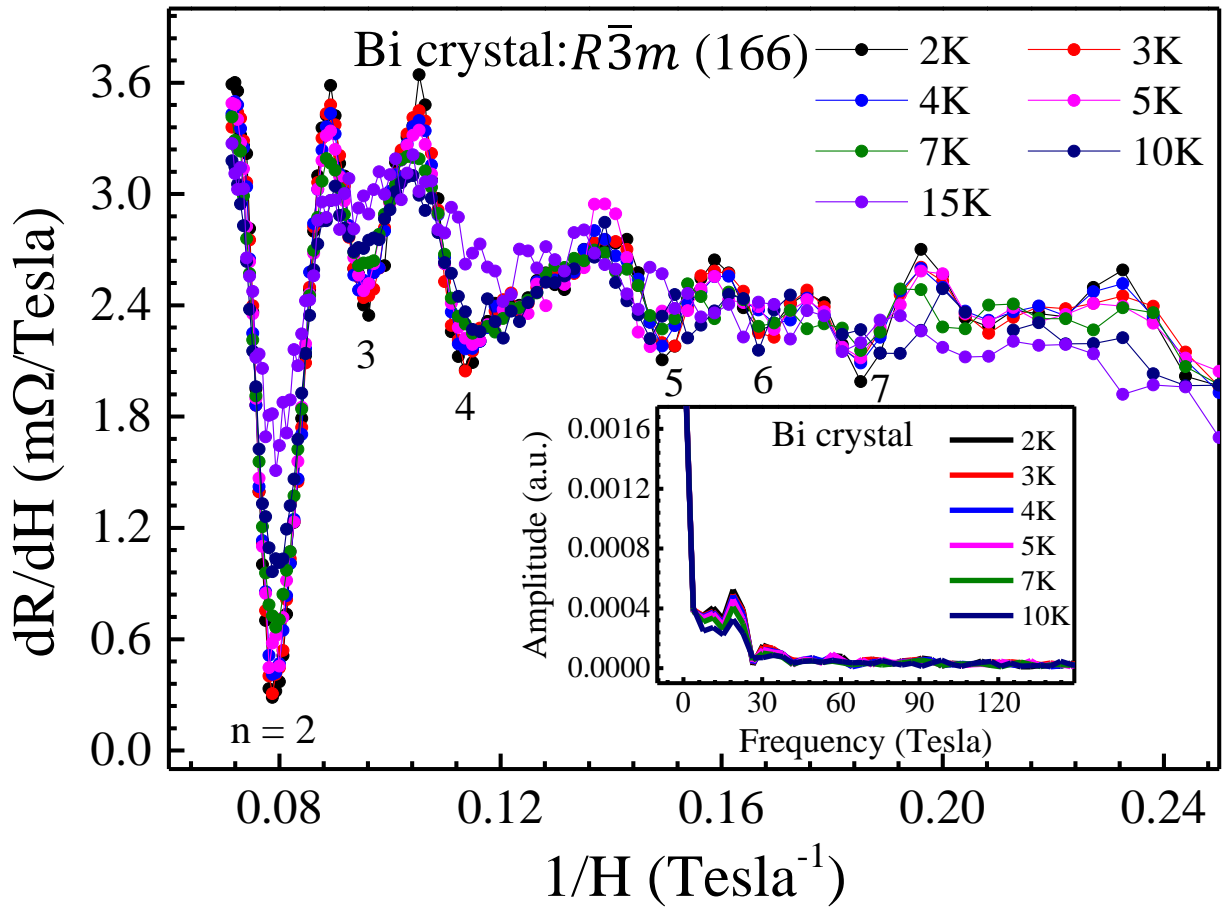

Fig. 5(b)

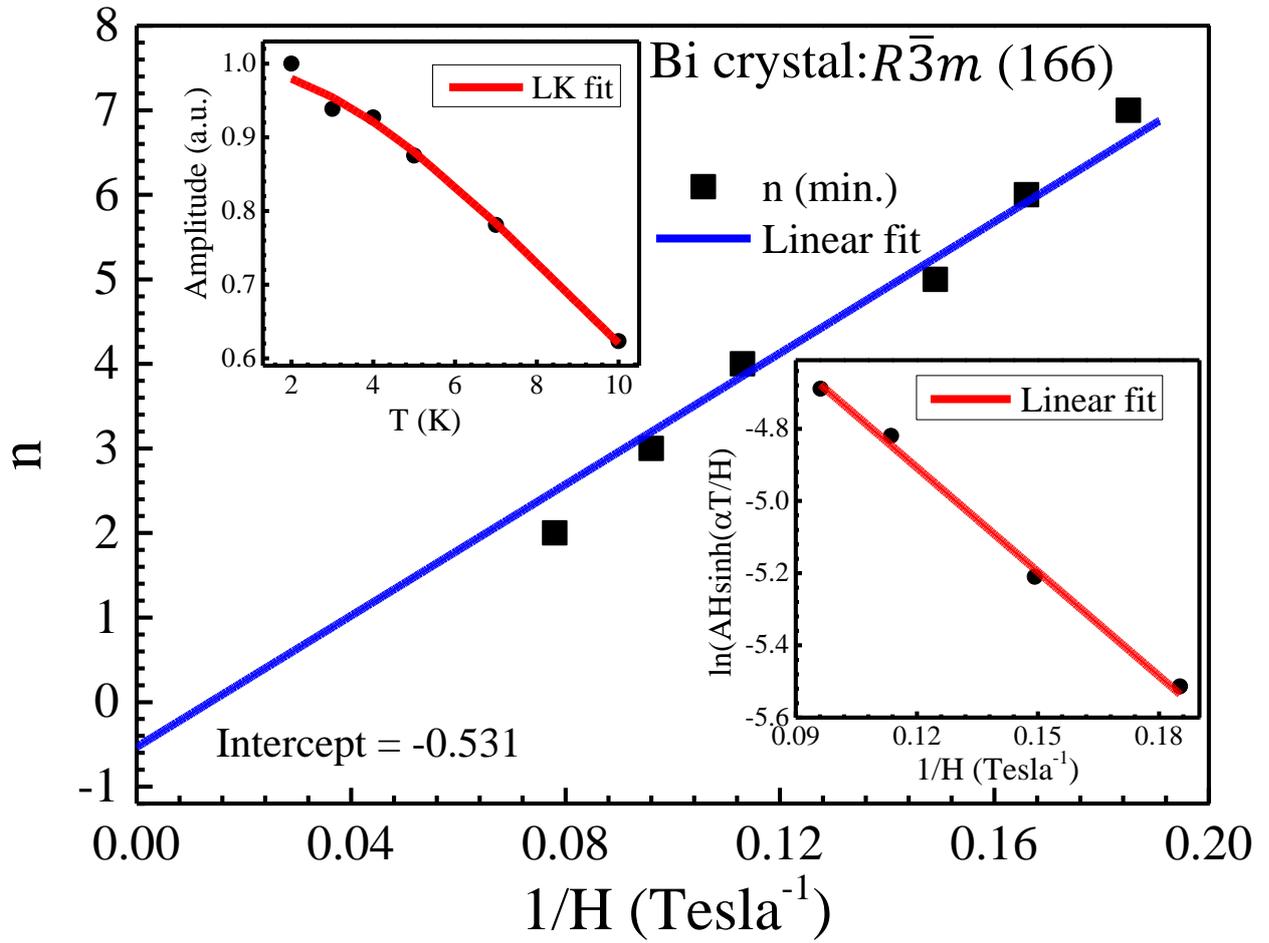